\documentclass[11pt]{article}
\usepackage{vietnam,epsfig}

\def\be{\begin{equation}}
\def\ee{\end{equation}}
\def\bea{\begin{eqnarray}}
\def\eea{\end{eqnarray}}

\begin{document}
\vspace*{4cm}
\title{GENERALIZED PARTON DISTRIBUTIONS: A NEW AVENUE 
TO COLOR TRANSPARENCY PHENOMENA}

\author{ S. LIUTI and S.K. TANEJA}

\address{Department of Physics, University of Virginia, 382 McCormick Road,\\
Charlottesville VA 22904, USA}

\maketitle\abstracts{
Color Transparency studies have been since long suggested 
as a means to study the occurrence and relevance of
small size hadronic configurations, predicted within QCD to dominate
exclusive scattering processes, by 
monitoring the passage of hadrons through the nuclear medium
at large four momentum transfer, $Q^2$.
The validation through experiments 
of this picture -- the dominance of short
separation components --  
is however not straightforward.
This situation motivated us to explore a description of Color Transparency
in terms of Generalized Parton Distributions (GPDs)
using a recent interpretation in impact parameter representation.}

\section{Introduction}
A most intensively studied question in Quantum ChromoDynamics (QCD)
is the space-time structure of high energy exclusive reactions. 
In the hard scattering approach
these are expected to be
dominated by the Fock space components of their wave function with the 
minimum number of quarks (anti-quarks). Such configurations, in turn, 
are only possible if the constituents are located 
within a small relative transverse 
distance, $ \approx 1/\sqrt{Q^2}$, 
$Q^2$ being the (high) four-momentum transfer squared in the 
reaction \cite{BroLep}. 

It was suggested in  Refs.\cite{BroMul,RalPir} 
that performing (quasi)-elastic reactions off nuclear targets 
can provide an experimental test of the dominance of short
separation components. Nuclei can in fact 
function as both ``passive'' or ``active'' probes for the small 
partonic separation components. At very large $Q^2$, 
small size configurations were in fact predicted to be 
less subject to rescatterings inside a nucleus with $A$ nucleons.
This is in turn a consequence of the fact that their cross 
section is expected to be  
proportional to their transverse size within 
the one gluon exchange QCD dipole model of high energy 
hadron-hadron scattering \cite{LowNus}.
Small distances can also be filtered at finite (moderate) $Q^2$, and 
varying $A$, by observing that large separations will gradually be blocked
by the strong interactions occurring in the nucleus, as $A$ increases. 

From a practical point of view, however,  
current searches for CT
might appear to be in a stall as 
all experiments performed so far seem not to show either any 
systematics or any marked trend for the onset of 
this phenomenon. 
Additional observables and new experiments have been recently proposed 
in order to better interpret the present situation, at the light, also, of  
emerging critical observations that the transverse size
of exclusive hard processes might not be small due to the persistence
of large endpoint contributions of the hadron's wave function \cite{Hoyer}.

In summary, whether or not a pQCD description of hadrons holds at the 
$Q^2$ values presently available, or at reach at future experimental 
programs, it has now become imperative to investigate the basic question
of the existence and observability of small size hadronic 
configurations. 
Generalized Parton Distributions (GPDs) have been recently shown 
to provide a theoretical tool for studying the 
(deeply inelastic) spatial structure of hadrons \cite{Bur,Die}.
Color Transparency and Nuclear Filtering are aimed at providing
measurements of the spatial extension of different hadronic components.
The usage of these two tools in combination represent a promising  
whole new dimension in studies of hadronic structure.

\section{Quantitative study of the transverse structure of the proton} 

We present here the results of a comprehensive study of the transverse
size of the proton, using GPDs. GPDs were introduced 
a few years ago with the main aim of providing a framework 
to describe in a partonic language 
the concept of orbital angular momentum carried by the nucleon's 
constituents \cite{DMul1,Ji1,Rad1}. 
In the deeply virtual Compton scattering reaction $ep \rightarrow e^\prime p \gamma$,
where the final photon is emitted from the proton's blob,  
one can describe the soft part of the reaction by introducing 
two GPDs, $H, E$, corresponding to the two possibilities or the 
final particle's helicity. 
The relevant kinematical variables in the process are: 
$P$ and $P^\prime$, the initial 
and final nucleon's momenta in the exclusive process,  
$\bar{P}=(P+P^\prime)/2$, the average nucleon momentum, $k$ the 
active quark momentum, $q$ the virtual photon momentum, $\Delta=P-P^\prime$ 
from which one obtains the relativistic invariants for the process: 
$x=k^+/\bar{P}^+$, $\xi=-\Delta^+/2\bar{P}^+$, $Q^2 = -q^2$, and
$t \equiv \Delta^2$ (for a review see {\it e.g.} Ref.\cite{Goe}).

More recently \cite{Bur}, a relationship was found between
GPDs and the Impact Parameter dependent Parton Distributions 
(IPPDs) defined as the joint distribution:
$\displaystyle d n/dx d{\bf b} \equiv q(x,{\bf b})$ --
the number of partons of type $q$ 
with momentum fraction $x=k^+/P^+$, located at
a transverse distance ${\bf b}$ (${\bf b}$ is the impact parameter) 
from the center of $P^+$ of the system.
\cite{Soper}. The 
connection is obtained by observing that for a purely 
transverse four momentum transfer, namely for 
$\Delta \equiv (\Delta_0=0; {\bf \Delta}, \Delta_3=0)$ and $\xi=0$,
$H_q(x,0,-{\bf \Delta}^2)$, 
and $q(x,{\bf b})$ can be 
related as follows:
\begin{eqnarray}
q(x,{\bf b}) & = & \int \frac{d^2 {\bf \Delta}}{(2 \pi)^2} \,
e^{-i {\bf b} \cdot {\bf \Delta}} H_q(x,0,-{\bf \Delta}^2)
\label{bdis1}  \\
H_q(x,0,-{\bf \Delta}^2) & = & \int d^2 {\bf b} \,
e^{i {\bf b} \cdot {\bf \Delta}} q(x,{\bf b}) .
\label{bdis2}  
\end{eqnarray}
Since $q(x,{\bf b})$ satisfies positivity constraints and it can
be interpreted as a probability distribution, $H_q(x,0,-{\bf \Delta}^2)$
is also interpreted as a probability distribution, namely the Fourier 
transformed joint probability distribution 
of finding a parton $i$ in the proton with longitudinal momentum fraction 
$x$, at the transverse position ${\bf b}$, with respect to the
center of momentum of the nucleon.

As shown in \cite{LiuTan_1}, the {\em radius} of the system of partons, 
which is needed for quantitative CT studies, is:
\begin{equation}
\displaystyle \, \langle {\bf r}^2(x) \, \rangle ^{1/2} = 
MAX\left\{ \langle {\bf b}^2(x) \rangle ^{1/2}, \langle {\bf b}^2(x) \rangle ^{1/2}
\frac{x}{1-x}   \right\}
\label{r}
\end{equation} 
In what follows, we describe the behavior of: 
{\it i)} 
the hadronic configuration's radius, $\langle {\bf r}^2(x) \rangle^{1/2}$;
{\it ii)} the intrinsic transverse momentum, ${\bf k}$;
{\it iii)} 
the average value of $x$ in elastic scattering from the proton: 
%%%%%
\begin{eqnarray}
\displaystyle
\langle \, {\bf b}^2(x) \, \rangle & = & {\cal N}_b \int d^2 {\bf b} \; q(x,{\bf b})
\, {\bf b}^2, \label{b}
\\
\langle \, {\bf k}^2(x) \, \rangle & = & {\cal N}_k \int d^2 {\bf k} \; f(x,{\bf k})
\, {\bf k}^2,
\label{k}
\\
\langle x(\Delta) \rangle & = & {\cal N}_x \int_0^1 dx \, x \, H(x,\Delta)
\label{x}
\end{eqnarray}
where ${\cal N}_b$, ${\cal N}_k$ and ${\cal N}_x$ are normalization factors, and
$\Delta=\sqrt{-t}$.
Furthermore, $\langle {\bf r}^2(x) \, \rangle ^{1/2}$ is obtained inserting 
Eq.(\ref{b}) in Eq.(\ref{r}); $f(x,{\bf k})$, the Unintegrated Parton Distribution
(UPD), is defined as: 
\begin{equation}   
f(x,{\bf k})  =  \int d^2 {\bf b} \int d^2 {\bf b}^\prime \, 
e^{i {\bf k} \cdot ({\bf b} - {\bf b}^\prime)} \, q(x,{\bf b}, {\bf b}^\prime),
\label{fk}
\end{equation}
where $q(x,{\bf b}, {\bf b}^\prime)$ is the non-diagonal IPDF. Notice that
$\langle x \rangle$ is calculated directly in terms of the GPD, $H$.

We compare three different models (see also \cite{LiuTan_1}), respectively 
characterized by:
{\bf (a)} a soft gaussian type distribution for $q(x,{\bf b})$ \cite{Rad98};
{\bf (b)} the ``semi-hard'' large $x$ model of \cite{Bur_new};
{\bf (c)} a quark-diquark model, characterized by
large intrinsic transverse momentum components $\propto 1/k^4$. 

\begin{figure}
\rule{5cm}{0.2mm}\hfill\rule{5cm}{0.2mm}
\vskip 2.5cm
%\rule{5cm}{0.2mm}\hfill\rule{5cm}{0.2mm}
\includegraphics[width=5.0cm]{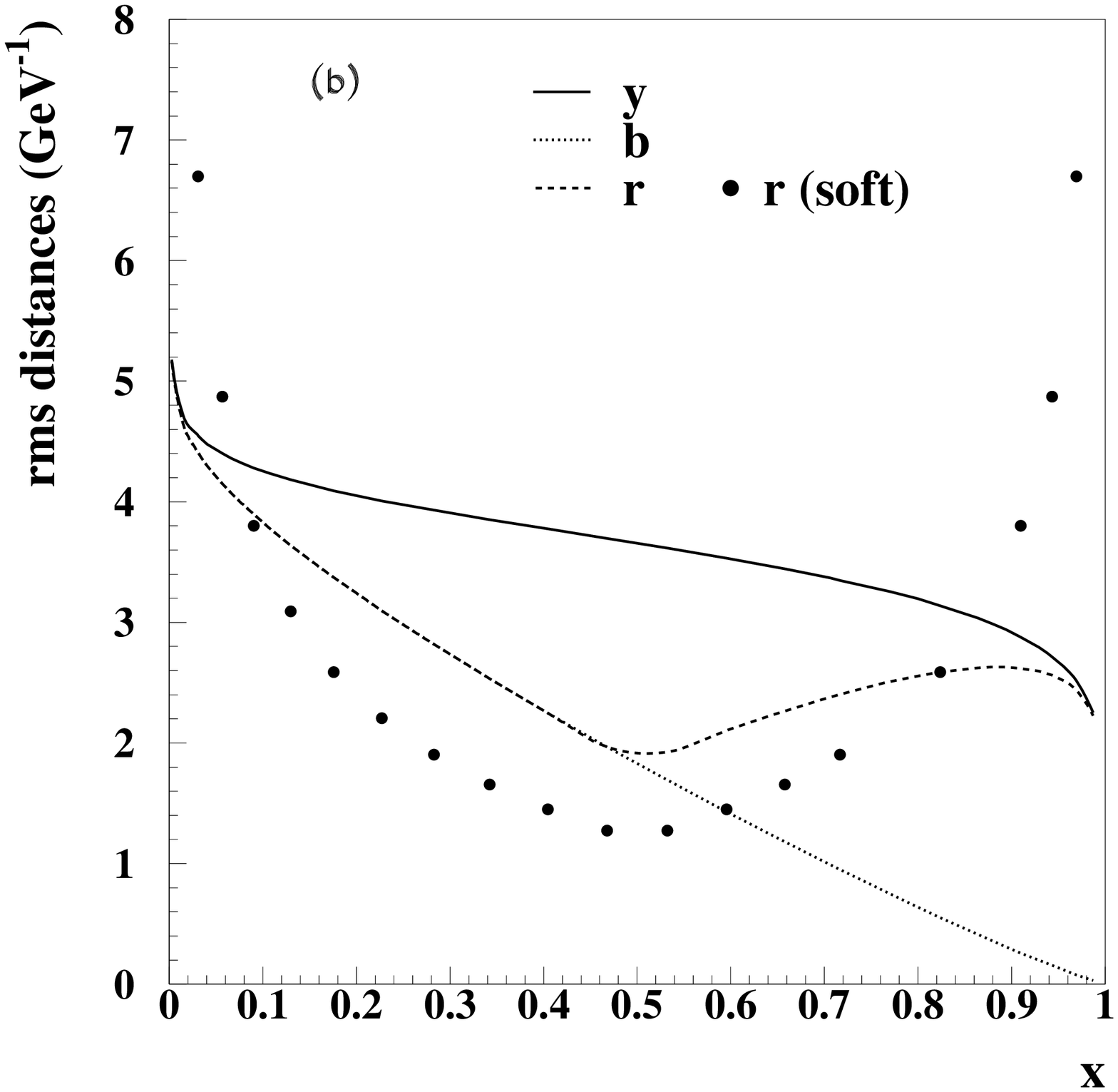}
\includegraphics[width=5.0cm]{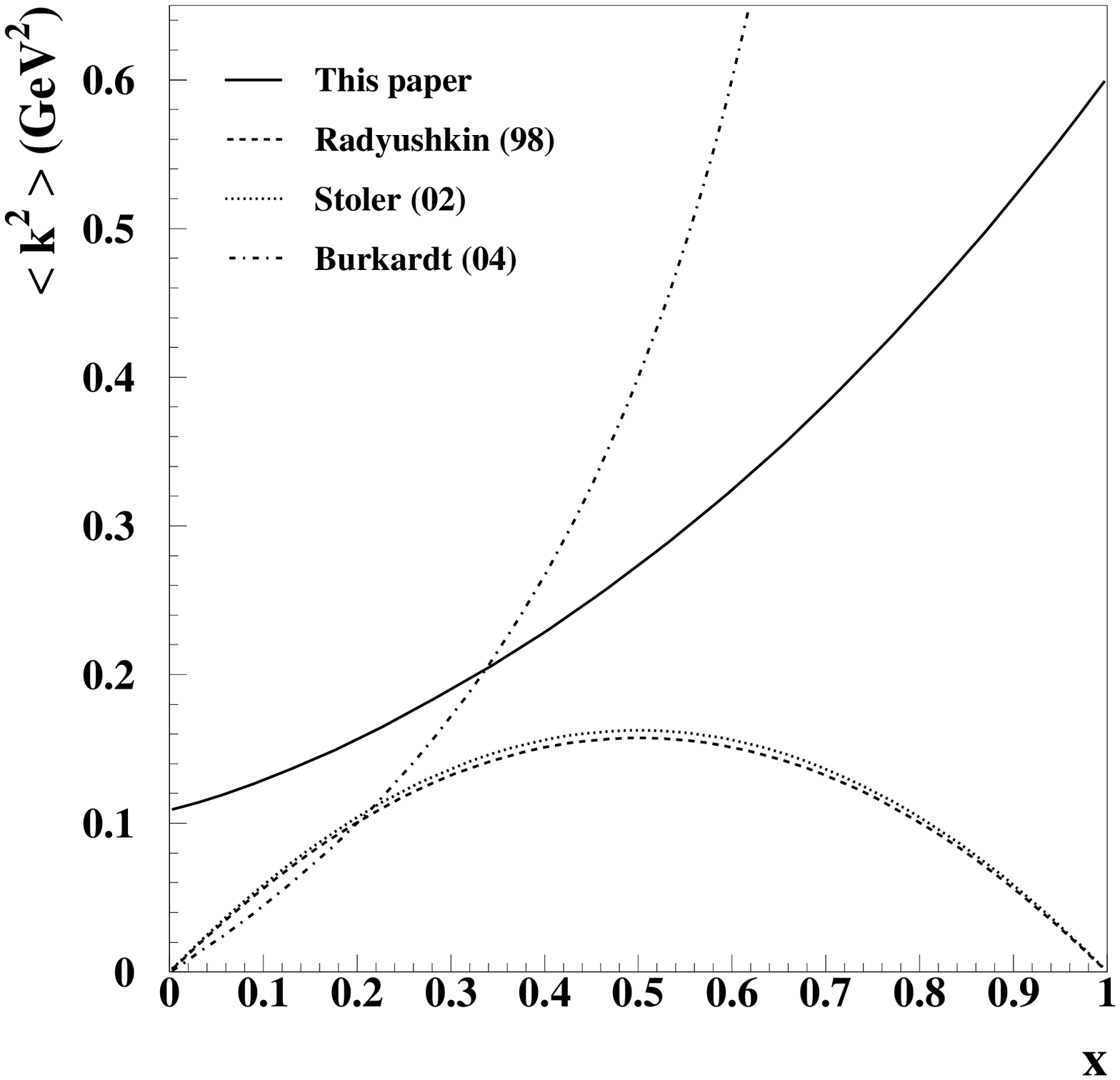}
\includegraphics[width=5.0cm]{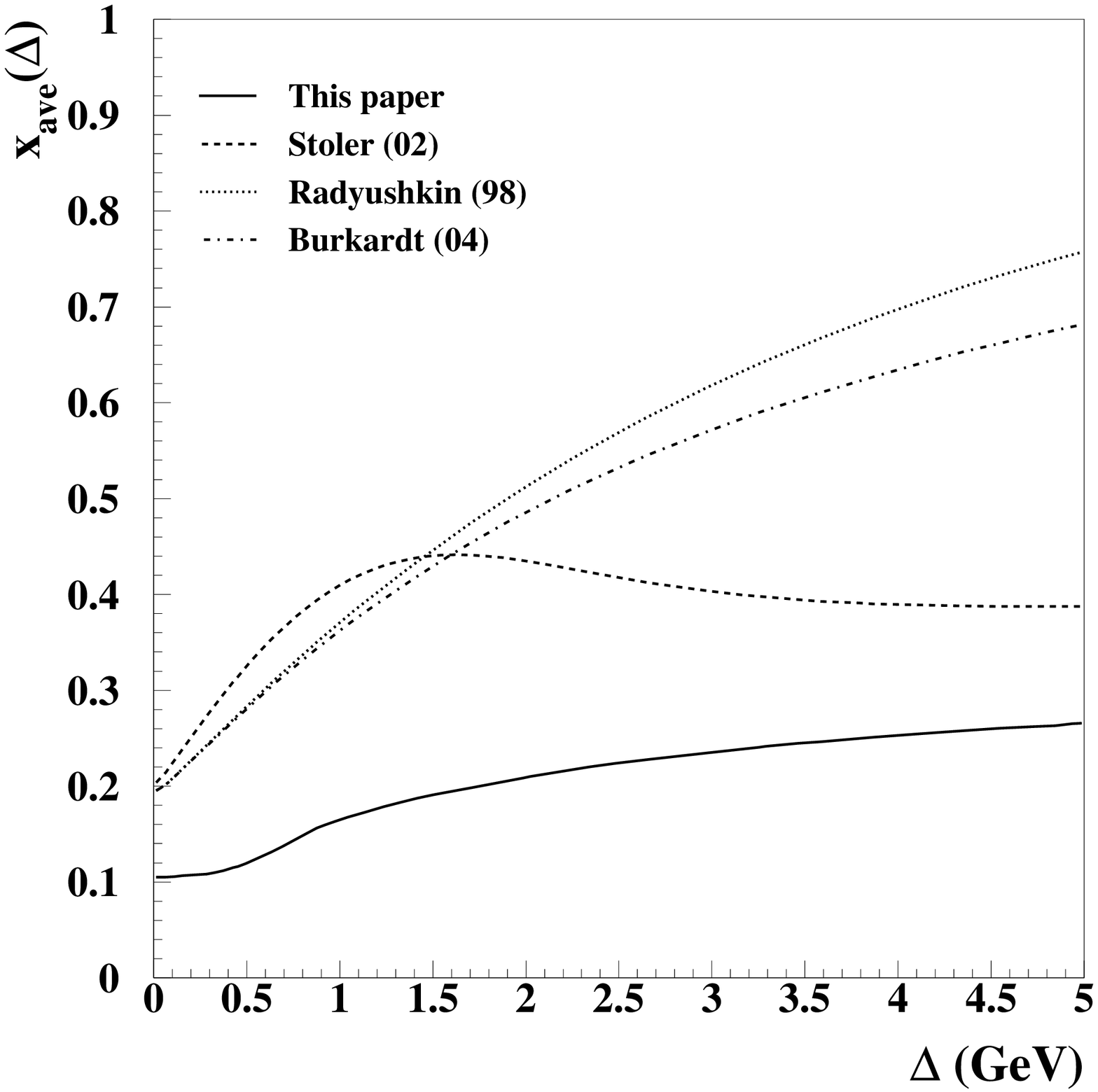}
\caption{The hadronic configuration's radius, Eqs.(\ref{r}) and (\ref{b}), (left);
the intrinsic transverse momentum, Eqs.(\ref{k}) and (\ref{fk}), (center);
and the average value of $x$, as a function of $\Delta=sqrt(-t)$, Eq.(\ref{x}).
\label{fig1}}
\end{figure}
  
Results are presented in Fig.\ref{fig1}. One can see that while the ``soft'' model
of Ref.\cite{Rad98} predicts an unphysically large value for the proton's radius 
at large $x$ (large dots in left panel), it is however not dominated by large
$x$ components at large momentum transfer (dotted curve in right panel).
On the other side, the model of Ref.\cite{Bur_new} gives a 
vanishing value 
of the radius at large $x$ (not shown in this figure), it is dominated by
large $x$ components at large $\Delta$ (dot-dashed curve in right panel), 
at the expense, however, of introducing very large intrinsic momentum 
components (dot-dashed curve in central panel). Finally, the quark-diquark model
stands in between the previous two models in that it predicts physically acceptable
{\em although non vanishing} behaviors for both the radius and the 
intrinsic ${\bf k}$ at large $x$; at the same time it is not dominated by large $x$
components at large momentum transfer.        

In summary, while it can be challenging to unambiguously disentangle the 
amount and nature of hard components responsible for the large $t$ behavior
of the hadronic form factors, one might 
gain a better insight by requiring models to simultaneously 
describe the hadrons transverse spatial distribution, and in particular
the possible onset of small transverse configurations.   
The diquark model presented here seems to 
provide a satisfactory starting point for such studies.

%%%
\begin{figure}
\vskip 2.5cm
\includegraphics[width=8.cm]{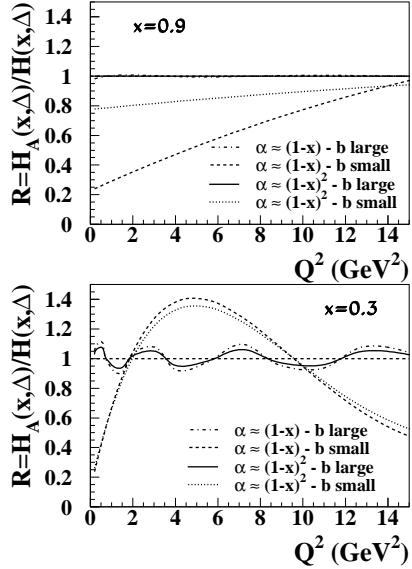}
\caption{Possible scenarios for the effect of nuclear filtering and 
the onset of CT based on the
models outlined in the text.}
\label{fig13}
\end{figure}
%%%
If configurations with small radii indeed exist,
they can be isolated in principle by performing CT and/or nuclear filtering
type experiments \cite{LiuTan_1,BurMil}. 
By considering $(e, e^\prime p)$ processes with an ``unmodified'' 
proton in the nuclear medium one has \cite{LiuTan_1}:  
\begin{equation}
H_A(t) = \int_0^{b_{max}(A)} db \, b \, q(x,b) J_0(b\Delta),
\label{transp1}
\end{equation}
where we introduced a nuclear filter for the 
large transverse size components 
by multiplying the IPDF, $q(x,b)$, by a square function:
\[ \Pi(b) = \left \{  \begin{array}{c} 
1  \; \; b <    b_{max}(A) \\
0 \; \;  b \geq b_{max}(A) \end{array} \right. ,  \]
%%%
$b_{max}(A)$ being the size of 
the filter. 
The transparency ratio is then defined as: 
\begin{equation}
\displaystyle T_A(Q^2) =  
\frac{\left[ \int_0^1 dx H_A(x,\Delta) \right] ^2}{\left[ \int_0^1 dx H(x,\Delta) \right]^2}, 
\end{equation}
where scattering in free 
space is calculated setting $b_{max}=\infty$ in the denominator.
Based on this result, one can fit the available data, using 
different distributions $q(x,b)$, and
varying the parameter $b_{max}$.   
In Fig.\ref{fig13} we compare: {\bf (a)} a soft distribution \cite{Rad98}, 
characterized by a parameter $\alpha \approx (1-x)$, with 
{\bf (b)} the harder ones proposed {\it e.g. } in \cite{LiuTan_1,Bur_new},
characterized by $\alpha \approx (1-x)^2$ (see \cite{LiuTan_1} for details).
The effect of the filter is to produce both damping and oscillations
in $H_A$. In Fig.\ref{fig13} we show for illustration, the ratio
$R = H_A(x,\Delta)/H(x,\Delta)$ plotted vs. $\Delta$ for two different values
of $x$, in case {\bf (a)} and {\bf (b)}, and
for different values of the filter size, ${\bf b}_{max}$. 
The results shown in the figure 
allow us to understand for varying $x$, the different effects due 
to variations 
in the size of $\alpha$, which in turn is a feature of different
models of GPDs. 
For instance, in the upper panel, $x=0.9$, for large ${\bf b} \equiv {b}_{max}$, 
CT is attained, independently from the model. For small ${\bf b} \equiv {b}_{max}$,
the soft model (case {\bf (a)}, \cite{Rad98}), is highly attenuated with respect
to the hard one (case {\bf (b)}, \cite{Bur_new,LiuTan_1}).

In conclusion our study of CT using the new concept of GPDs, 
will both improve our knowledge of 
nuclear filtering phenomena  
and allow for a more detailed understanding of the transverse 
components involved at large momentum transfer.  
In particular, we hope to have provided a connection between
${\bf b}-$ and ${\bf k}-$ space that will help 
to systematically address both the role 
of Sudakov effects in
the endpoints of the hadronic wave function \cite{Hoyer}, 
and the role of power corrections in the large longitudinal
momentum region.

\section*{Acknowledgments}
This work is supported by the U.S. Department
of Energy grant no. DE-FG02-01ER41200.

\end{document}